\documentclass[draft,twoside]{article}
\newcommand{\mytitle}{Ensemble Averages when
  $\beta$ is a Square Integer} 
\newcommand{\keywords}{Random matrix theory, partition function,
  Pfaffian, hyperpfaffian, Selberg integral}

\usepackage{amsmath,amssymb, amsthm,amsfonts,amscd,mathrsfs,pifont} 
\usepackage{eucal}  
\usepackage{verbatim}     
\usepackage[dvips]{graphicx}

\newtheorem{thm}{Theorem}[section]

\newtheorem{thm*}{Theorem}[]
\newtheorem{cor*}[thm*]{Corollary}
\newtheorem{claim*}[thm*]{Claim}
\newtheorem{lemma*}[thm*]{Lemma}
\newtheorem{prop*}[thm*]{Proposition}
\newtheorem{conj*}[thm*]{Conjecture}

\theoremstyle{definition}

\newtheorem{question*}{Question}[section]

\newtheorem{defn*}{Definition}

\theoremstyle{remark}

\newcommand{\BB}[1]{\ensuremath{\mathbb{#1}}}

\newcommand{\R}{\ensuremath{\BB{R}}}

\newcommand{\C}{\ensuremath{\BB{C}}}

\newcommand{\bs}{\ensuremath{\boldsymbol}}
\newcommand{\mf}{\ensuremath{\mathfrak}}

\newcommand{\wt}{\ensuremath{\widetilde}}

\newcommand{\qand}{\qquad \mbox{and} \qquad}

\newcommand{\qthen}{\qquad \mbox{then} \qquad}

\newcommand{\ul}[1]{\underline{#1}}

\newcommand{\G}[1]{\Gamma\left( #1 \right)}

\DeclareMathOperator{\sgn}{sgn}
\DeclareMathOperator{\vol}{vol}

\DeclareMathOperator{\Pf}{Pf}
\DeclareMathOperator{\PF}{PF}

\usepackage{ifpdf}
\ifpdf
\fi

\newcommand{\Wr}{\ensuremath{\mathrm{Wr}}}
\newcommand{\Vol}{\ensuremath{\mathrm{Vol}}}

\numberwithin{equation}{section} 

\numberwithin{equation}{section}

\begin{document}
\title{\bfseries\sffamily \mytitle}  
\author{\sc Christopher D.~Sinclair\footnote{This research was 
    supported in part by the National Science Foundation (DMS-0801243)} }
\maketitle

\begin{abstract}
We give a hyperpfaffian formulation of partition functions and
ensemble averages for Hermitian and circular ensembles when $L$ is an
arbitrary integer and $\beta = L^2$ and when $L$ is an odd integer and
$\beta = L^2 + 1$.   
\end{abstract}

\noindent {\bf Keywords:} \keywords
\vspace{1cm}

For many ensembles of random matrices, the joint probability density of
eigenvalues has the form
\begin{equation}
\label{eq:8}
\Omega_N(\bs \lambda) =  \frac{1}{Z_N N!} \bigg\{ \prod_{n=1}^N
w(\lambda_n) \bigg\}
\prod_{m<n} | \lambda_n - \lambda_m |^{\beta}, \qquad \bs \lambda \in
\R^N 
\end{equation}
where, $\beta > 0$, $w: \R \rightarrow [0,\infty)$ is a weight
function and $Z_N$ is a normalization constant called the {\em
  partition function}. The parameter $\beta$ is often taken to be 1,2
or 4, since for these values the correlation functions ({\em
  viz.}~marginal probabilities) have a determinantal or 
Pfaffian form.

The partition function is given explicitly by 
\begin{equation}
\label{eq:2}
Z_N = Z_N^{\nu} = \frac{1}{N!} \int_{\R^N} \prod_{m<n} | \lambda_n - \lambda_m |^{\beta}
\, d\nu^N(\bs \lambda),
\end{equation}
where $d\nu(\lambda) = w(\lambda) d\lambda$ and $\nu^N$ is the
resulting product measure on $\R^N$.  However, for now, we will take
$\nu$ to be arbitrary, with the restriction that $0 < Z_N < \infty$,
and we will suppress the notational dependence of $Z_N$ on $\nu$
except when necessary to allay confusion.  Ensemble averages of
functions which are multiplicative in the eigenvalues can be expressed
as integrals of the form (\ref{eq:2}) where $\nu$ is determined by the
function on the eigenvalues, the weight function of the ensemble and
the underlying reference measure.

In the case where $\beta = 2\gamma$ and 
\[
d\nu(\lambda) = \lambda^{a-1} (1- \lambda)^{b - 1}
\bs{1}_{[0,1]}(\lambda) \, d\lambda; \hspace{1cm} a, b, \gamma \in \C,
\]
$S_N(\gamma, a, b) = N! Z_N$ is the (now) famous Selburg integral \cite{MR0018287,
  MR1117906}, and in this case
\[
S_N(\gamma, a ,b) = \prod_{n=0}^{N-1} \frac{\G{a + n \gamma} \G{b + n
    \gamma} \G{1 +(n+1)\gamma}}{\G{a + b + (N+n-1)\gamma} \G{1 + \gamma}}.
\]
The case where
\[
d\nu(\lambda) = \frac{1}{\sqrt{2 \pi}} e^{-\lambda^2/2} \qand \beta =
2 \gamma,
\]
$F_N(\gamma) = N! Z_N$ is called the Mehta integral \cite{MR1190440,
  MR0151232}, and has the evaluation
\[
F_N(\gamma) = \prod_{n=1}^{N} \frac{\G{1 + n \gamma}}{\G{1 + \gamma}}.
\]
This evaluation follows the same basic idea as that of the Selberg
integral.  For an interesting history of these evaluations, see
\cite{MR2434345}.  

The purpose of this note is not to provide further explicit
evaluations of for particular choices of $\nu$, but rather to show
that when $\beta$ is a square integer (or adjacent to a square
integer) that $Z_N$ can be expressed as a hyperpfaffian.  Expressing
ensemble averages as a determinant or Pfaffian is the first step to
demonstrating the solvability of an ensemble; that is writing the
correlation functions, gap probabilities and other quantities of
interest in terms of determinants or Pfaffians formed from a kernel
associated to the particulars of the ensemble
\cite{MR1657844, sinclair-2008, borodin-2008}.  Pfaffian formulations
for $Z_N$ are known when $\beta=1$ and $\beta=4$, and the results here
generalize those formulas.  To be specific, we show that when $L$ is
odd and $\beta = L^2$ or $L^2 + 1$, then $Z_N$ can be written as the
hyperpfaffian of a $2L$-form, the coefficients of which are formed
from double integrals of products of $L \times L$ Wronskians of monic
polynomials in a manner which generalizes the well known $\beta=1$ and
$\beta=4$ cases.  When $L$ is even, $Z_N$ can be written as a
hyperpfaffian of an $L$-form, the coefficients of which are integrals
(with respect to the measure $\nu$) of Wronskians of any complete
family of monic polynomials. This generalizes the situation when
$\beta = 4$.  Similar results for $L$ even were given in
\cite{MR1943369}.

We will also give a similar characterization of ensemble averages over
circular ensembles.  

\section{Pfaffians, Hyperpfaffians and Wronskians}

\subsection{Pfaffians and Hyperpfaffians}
Suppose $V = \R^{NL}$ with basis $\mathbf e_1, \mathbf e_2, \ldots, \mathbf
e_{NL}$.  For each increasing function $\mf t$, from $\ul L =
\{1,2,\ldots, L\}$ into $\ul{NL} = \{1,2, \ldots, NL\}$, we write 
\[
\epsilon_{\mf t} = \mathbf e_{\mf t(1)} \wedge \mathbf e_{\mf t(2)}
\wedge \cdots \wedge \mathbf e_{\mf t(L)},
\]
so that $\{ \epsilon_{\mf t} \; \big| \; \mf t: \ul L \nearrow \ul{NL}
\}$ is a basis for $\Lambda^L V$.  The volume form of $\R^{NL}$ is
given by   
\[
\epsilon_{\Vol} = \mathbf e_1 \wedge \mathbf e_2 \wedge \cdots \wedge
\mathbf e_{NL}.
\]
The $N$-fold wedge product of an $L$-form $\omega$ is a constant times
$\epsilon_{\Vol}$. This constant is (up to a factor of $N!$) the {\em
  hyperpfaffian} of $\omega$.  Specifically, for $\omega \in \Lambda^L
V$, we define $\PF \omega$ by  
\[
\frac{\omega^{\wedge N}}{N!} = \frac{1}{N!}\underbrace{\omega \wedge
  \omega \wedge \cdots   \wedge \omega}_N =\PF \omega \cdot
\epsilon_{\Vol}. 
\]

When $L=2$, the hyperpfaffian generalizes the notion of the Pfaffian
of an antisymmetric matrix:  If $\mathbf A = \left[ a_{m,n} 
\right]$ is an antisymmetric $2N \times 2N$ matrix, then the Pfaffian
of $\mathbf A$ is given by 
\[
\Pf \mathbf A = \frac{1}{2^N N!} \sum_{\sigma \in S_{2N}} \sgn \sigma
\prod_{n=1}^N a_{\sigma(2n-1),\sigma(2n)}.
\]
We may identify $\mathbf A$ with the 2-form  
\[
\alpha = \sum_{m < n} a_{m,n} \mathbf e_{m} \wedge \mathbf e_n.
\]
It is an easy exercise to show that $\Pf \mathbf A = \PF \alpha$.
There exists a formula for the hyperpfaffian as a sum over the
symmetric group, and in fact, the original definition
of the hyperpfaffian was given as such a sum \cite{MR1943369,
  MR1355700}. 

\subsection{Wronskians}

A {\em complete} family of monic polynomials is an $NL$-tuple of
polynomials $\mathbf p = \left( p_n \right)_{n=1}^{NL}$ such that
each $p_n$ is monic and $\deg p_n = n-1$.   Given an increasing
function $\mf t: \ul L \nearrow \ul{NL}$, we define the $L$-tuple
$\mathbf p_{\mf t} = \left( p_{\mf t(\ell)} \right)_{\ell=1}^L$.  And,
given $0 \leq \ell < L$ we define the modified $\ell$th
differentiation operator by 
\begin{equation}
\label{eq:1}
D^{\ell}f(x) = \frac{1}{\ell!} \frac{d^{\ell}f}{dx^{\ell}}
\end{equation}
The
{\em Wronskian} of $\mathbf p_{\mf t}$ is then defined to be
\[
\Wr(\mathbf p_{\mf t}; x) = \det \left[ D^{\ell-1} p_{\mf t(n)}(x)
\right]_{n,\ell=1}^L. 
\]
The Wronskian is often defined without the $\ell!$ in the denominator
of~(\ref{eq:1}); this combinatorial factor will prove convenient in
the sequel.  The reader has likely seen Wronskians in elementary
differential equations, where they are used to test for linear
dependence of solutions.   

\section{Statement of Results}

For each $x \in \R$, we define the $L$-form $\omega(x) \in \Lambda^L
V$ by
\begin{equation}
\label{eq:4}
\omega(x) = \sum_{\mf t: \ul L \nearrow \ul{NL}} \Wr(\mathbf p_{\mf
  t}; x) \epsilon_{\mf t}.
\end{equation}
This form clearly depends on the choice of $\mathbf p$, though we will
suppress this dependence.  We may arrive at an 
$L$-form with constant coefficients by integrating the
coefficients of the form with respect to $d\nu$.  In this situation we
will write
\[
\int_{\R} \omega(x) \, d\nu(x) = \sum_{\mf t: \ul{L} \nearrow \ul{NL}}
\bigg\{ \int_{\R} \Wr(\mathbf p_{\mf t}; x) d\nu(x) \bigg\}
\epsilon_{\mf t}.
\]
Notice that we are not integrating the form in the sense of
integration on manifolds, but rather formally extending the linearity
of the integral over the coefficient functions of $\omega(x)$ to arrive at
an $L$-form with constant coefficients.  

To deal with the $N$ odd case, we set $V' =
\R^{(N+1) L}$ and define the basic $L$-form $\epsilon' = \mathbf
e_{NL+1} \wedge \mathbf e_{NL+2} \wedge \cdots \wedge \mathbf e_{NL+L}$. 
\begin{thm}
\label{thm:1}
Suppose $L$ and $N$ are positive integers, and let
$\omega(x)$ be the $L$-form given as in (\ref{eq:4}) for any complete
family of monic polynomials $\mathbf p$.  Then, 
\begin{enumerate}
\item\label{item:1} if $\beta = L^2$ is even, 
\[
Z_N = \PF \bigg( \int_{\R} \omega(x) \, d\nu(x) \bigg);
\]
\item if $\beta = L^2$ is odd and $N$ is even, 
\[
Z_N = \PF \bigg(\frac12 \int_{\R} \int_{\R} \omega(x) \wedge \omega(y)
\sgn(y-x) \, d\nu(x) \, d\nu(y) \bigg);
\]
\item \label{item:2} if $\beta = L^2$ is odd and $N$ is odd, 
\[
Z_N = \PF \bigg(\frac12 \int_{\R} \int_{\R} \omega(x) \wedge \omega(y) \sgn(y-x) \,
d\nu(x) \, d\nu(y) +
\int_{\R} \omega(x) \wedge \epsilon' \, d\nu(x) \bigg);
\]
\item if $\beta = L^2 + 1$ is even and $N = 2M$ is even, 
\[
Z_N = \PF \bigg(
\frac12 \int_{\R} \int_{\R} \omega(x) \wedge \omega(y) \left[\frac{(y^M
    - x^M)^2}{y - x} \right] \, d\nu(x) \, d\nu(y)
\bigg).
\]
\end{enumerate}
\end{thm}
More explicitly, when $\beta = L^2$ is even we have
\[
\int_{\R} \omega(x) \, d\nu(x) = \sum_{\mf t: \ul L \nearrow \ul{NL}}
\int_{\R} \Wr(\mathbf p_{\mf t}; x) \, d\nu(x) \epsilon_{\mf t}
\]
In the case where $\beta = 4$, we have 
\[
\Wr(p_{\mf t}; x) = p_{\mf t(1)}(x) p'_{\mf t(2)}(x) - p_{\mf
  t(1)}'(x) p_{\mf t(2)}(x); \qquad \mf t: \ul 2 \nearrow \ul{2N},
\]
and 
\[
\int_{\R} \omega(x) \, d\nu(x) = \sum_{\mf t: \ul 2 \nearrow \ul{2N}} \bigg\{ \int_{\R} \big[ p_{\mf t(1)}(x) p'_{\mf 
  t(2)}(x) - p_{\mf   t(1)}'(x) p_{\mf t(2)}(x) \big] d\nu(x) \bigg\}
\epsilon_{\mf t(1)} \wedge \epsilon_{\mf t(2)}.
\]
This is exactly the $2$-form associated to the antisymmetric matrix
\[
\mathbf W = \left[  \int_{\R} \big[ p_{m}(x) p'_{n}(x) - p_{m}'(x) p_{n}(x) \big] d\nu(x) \right]_{m,n=1}^{2N},
\]
and $Z_N = \Pf \mathbf W$, as is well known \cite{MR0079647, MR2129906}. 

When $\beta = L^2$ is odd and $N$ is even, 
\begin{align*}
& \frac12 \int_{\R} \int_{\R} \omega(x) \wedge \omega(y) \sgn(y-x) \,
d\nu(x) \, d\nu(y) \\
&\hspace{2cm}= \sum_{\mf t, \mf u: \ul L \nearrow \ul{NL}}
\bigg\{\frac12  \int_{\R} \int_{\R} \Wr(\mathbf p_{\mf t}; x)
\Wr(\mathbf p_{\mf u}; y) \, \sgn(y-x) \, d\nu(x) \, d\nu(y) \bigg\}
\epsilon_{\mf t} \wedge \epsilon_{\mf u}
\end{align*}
In the case where $L = \beta = 1$, each $\mf t: \ul 1 \nearrow
\ul N$ selects a single integer between 1 and $N$, and 
$\Wr(\mathbf
p_{\mf t}; x) = p_{\mf t(1)}(x)$.  It follows that   \begin{align*}
& \frac12 \int_{\R} \int_{\R} \omega(x) \wedge \omega(y) \sgn(y-x) \,
d\nu(x) \, d\nu(y) \\
& \hspace{3cm} =  \sum_{m,n=1}^N \bigg\{ \frac12 \int_{\R}
\int_{\R} p_m(x) p_n(y) \sgn(y-x) \, d\nu(x) \, d\nu(y) \bigg\}
\mathbf e_m \wedge \mathbf e_n  
\end{align*}
This is the 2-form associated to the antisymmetric matrix
\[
\mathbf U = \left[ \frac{1}{2} \int_{\R} \int_{\R}  p_n(x) p_m(y)  \sgn(y-x)
\, d\nu(x) \, d\nu(y) \right]_{n,m=1}^N,
\]
and deBruijn's identities in this case have that $Z_N = \Pf \mathbf
U$. 

When $L = 1$ and $N$ is odd, the additional term on the right hand 
side of (\ref{item:2}) correspond to bordering the matrix $\mathbf U$ to produce
\[
\mathbf U' = \begin{bmatrix}
& & & \int_{\R} p_1(x) \, d\nu(x) \\
&  \mathbf U & & \vdots \\
& & & \int_{\R} p_N(x) \, d\nu(x) \\ 
-\int_{\R} p_1(x) \, d\nu(x) & \cdots & -\int_{\R} p_N(x) \, d\nu(x) & 0 
\end{bmatrix},
\]
and in this situation, $Z_N = \Pf \mathbf U'$ as is well known \cite{MR1762659}.  

The case where $\beta = L^2 + 1$ is even and $N = 2M$ does not seem to
generalize any known situation, though it is apparently applicable
when $\beta = 2$.  The partition functions of $\beta = 2$ ensembles
are traditionally described in terms of determinants.  And, while
every determinant can be written trivially as a Pfaffian, the Pfaffian
expression which appears for the partition function here does not seem
to arise from such a trivial modification.  At any rate, when
$\beta = 2$ we have that
\[
Z_{2M} = \Pf \mathbf Y,
\]
where
\begin{equation}
\label{eq:14}
\mathbf Y = \bigg[
\frac12 \int_{\R} \int_{\R} p_m(x) p_n(y) \bigg( \frac{(y^M -
    x^M)^2}{y -x} \bigg) \, d\nu(x) \, d\nu(y) 
\bigg]_{m,n=1}^{2M}.
\end{equation}

\subsection{For Circular Ensembles}

The proof of Theorem~\ref{thm:1} is mostly formal, and similar results
can be had for other ensembles.  For instance, the joint density of
Dyson's circular ensembles  \cite{MR0143556}  is given by
\[
Q_N(\bs \lambda) = \frac{1}{C_N N!} \prod_{m<n} | e^{i \theta_n} -
e^{i \theta_m}|^{\beta}; \qquad \theta_1, \theta_2, \ldots, \theta_N \in [-\pi,\pi)
\]
where
\[
C_N = \frac1{N!}\int\limits_{[-\pi,\pi)^N}\prod_{m<n} | e^{i\theta_n} - e^{i\theta_m}
|^{\beta} \,  d\theta_1 \, d\theta_2 \cdots d\theta_N.
\]

By \cite[(11.3.2)]{MR2129906}, 
\[
| e^{i\theta_n} - e^{i\theta_m} | = -i e^{-i(\theta_n
  + \theta_m)/2} \sgn(\theta_n - \theta_m) (e^{i\theta_n} - e^{i\theta_m}),
\]
and thus, if we define,
\begin{equation}
\label{eq:12}
d\mu(\theta) = (-ie^{-i \theta})^{(N-1)/2} \, d\theta,
\end{equation}
then
\[
C_N = \frac1{N!} \int\limits_{[-\pi,\pi)^N} \bigg\{ \prod_{m<n} ( e^{i\theta_n} -
e^{i\theta_m} ) \sgn(\theta_n - \theta_m) \bigg\}^{\beta} \,  d\mu^{N}(\bs \theta).
\]
Of course the $\sgn(\theta_n -
\theta_m)$ terms can be ignored when $\beta$ is even.  

An explicit evaluation of $C_N$ in the case where $d\mu(\theta) =
d\theta/{2\pi}$, conjectured in \cite{MR0143556} and proved by various
means in \cite{gunson:752, good:1884, askey:938, Zeilberger1982317}.

\begin{thm}
\label{thm:2}
Suppose $L$ and $N$ are positive integers with $\beta = L^2$ and let
$\omega(x)$ be the $L$-form given as in (\ref{eq:4}) and suppose $\mu$
is given as in (\ref{eq:12}).  Then, for any complete
family of monic polynomials $\mathbf p$,
\begin{enumerate}
\item if $\beta = L^2$
\[
C_N = \PF \bigg( \int_{-\pi}^{\pi} \omega(e^{i \theta}) \, d\mu(\theta) \bigg);
\]
\item if $\beta = L^2$ is odd and $N$ is even
\[
C_N = \PF \bigg( \frac12 \int_{-\pi}^{\pi} \int_{-\pi}^{\pi} \omega(e^{i \theta})
\wedge \omega(e^{i \psi}) \sgn(\psi-\theta) \, d\mu(\theta) \,
d\mu(\psi) \bigg); 
\]
\item if $\beta = L^2$ is odd and $N$ is odd
\[
C_N = \PF \bigg(\frac12 \int_{-\pi}^{\pi} \int_{-\pi}^{\pi}
\omega(e^{i \theta}) \wedge \omega(e^{i \psi})
\sgn(\psi-\theta) \, d\mu(\theta) \, d\mu(\psi) +
\int_{-\pi}^{\pi} \omega(e^{i \theta}) \wedge \epsilon' \,
d\mu(\theta) \bigg); 
\]
\item if $\beta = L^2 + 1$ is even and $N = 2M$ is even, 
\[
C_N = \PF \bigg(
\frac12 \int_{-\pi}^{\pi} \int_{-\pi}^{\pi} \omega(e^{i \theta}) \wedge \omega(e^{i
  \psi}) \left[\frac{(e^{i M \psi}  - e^{i M \theta})^2}{e^{i \psi} -
    e^{i \theta}} \right] \, d\mu(\theta) \, d\mu(\psi)
\bigg).
\]
\end{enumerate}
\end{thm}
The proof of this theorem is the same, {\it mutadis mutandis},
as that of Theorem~\ref{thm:1}.  

\subsection{In Terms of Moments}

For $j, k \geq 0$, let the $k$th moment of $\nu$ be given by
\[
M(k) = \int_{\R} x^k \, d\nu(x), 
\]
and the $j,k$th {\em skew}-moment of $\nu$ to be
\[
M(j,k) = \frac{1}{2} \int_{\R} \int_{\R} x^j y^k \sgn(y - x) \,
d\nu(x) \, d\nu(y).
\]
If we set $\mathbf p$ to be just the monomials, and define 
\[
\Delta\mf t = \prod_{j < k} \big(\mf t(k) - \mf t(j)\big) \qand
\Sigma\mf t = \sum_{\ell=1}^L \mf t(\ell),
\]
then the Wronskian of $\mathbf p_{\mf t}$ is given by \cite{dumas-bostan}
\[
\Wr(\mathbf p_{\mf t}; x) = \Delta\mf t \bigg\{\prod_{\ell=1}^L
\frac{1}{\big(\mf t(\ell) - 1\big)!} \bigg\}  x^{\Sigma
  \mf t - L(L+1)/2}.
\]
It follows that
\begin{align*}
\omega(x) &= \sum_{\mf t: \ul L \nearrow \ul{NL}}  \Delta \mf t
\bigg\{\prod_{\ell=1}^L \frac{1}{\big(\mf t(\ell) - 1\big)!} \bigg\}  \,
x^{\Sigma \mf t - L(L+1)/2} \epsilon_{\mf t}, \\
\int_{\R} \omega(x) \, d\nu(x) &= \sum_{\mf t: \ul L \nearrow \ul{NL}}
\Delta \mf t \bigg\{\prod_{\ell=1}^L \frac{1}{\big(\mf t(\ell) - 1\big)!} \bigg\}
 M\bigg( \Sigma \mf t -
  \frac{L(L+1)}{2}\bigg) \epsilon_{\mf t}  
\end{align*}
\begin{align*}
& \frac12 \int_{\R^2} \omega(x) \omega(y) \sgn(y-x) \, d\nu(x) \,
d\nu(y) \\
& \hspace{.5cm} = \sum_{\mf t, \mf u} \Delta \mf t \Delta \mf u
\bigg\{\prod_{\ell=1}^L \frac{1}{\big(\mf t(\ell) - 1\big)! \big(\mf
  u(\ell) - 1\big)!} \bigg\}  M\bigg( \Sigma \mf t -
  \frac{L(L+1)}{2}, \Sigma \mf u -
  \frac{L(L+1)}{2}\bigg) \epsilon_{\mf t} \wedge \epsilon_{\mf u}.
\end{align*}
Similarly, 
\begin{align*}
& \frac12 \int_{\R^2} \omega(x) \omega(y) \sgn(y-x) \, d\nu(x) \,
d\nu(y) + \int_{\R} \omega(x)\wedge\epsilon' \, d\nu(x) \\
& \hspace{.5cm} = \sum_{\mf t, \mf u} \Delta \mf t \Delta \mf u
\bigg\{\prod_{\ell=1}^L \frac{1}{\big(\mf t(\ell) - 1\big)! \big(\mf
  u(\ell) - 1\big)!} \bigg\}  M\bigg( \Sigma \mf t -
  \frac{L(L+1)}{2}, \Sigma \mf u -
  \frac{L(L+1)}{2}\bigg) \epsilon_{\mf t} \wedge \epsilon_{\mf u} \\
& \hspace{4.5cm} + \sum_{\mf t: \ul L \nearrow \ul{NL}}
\Delta \mf t \bigg\{\prod_{\ell=1}^L \frac{1}{\big(\mf t(\ell) - 1\big)!} \bigg\}
 M\bigg( \Sigma \mf t -
  \frac{L(L+1)}{2}\bigg) \epsilon_{\mf t}  \wedge \epsilon'.
\end{align*}

\section{A Remark on Correlation Functions}

We consider the joint density $\Omega_N(\bs \lambda)$ given as in
(\ref{eq:8}), though the results here easily extend to circular and
other related ensembles.  We will define the measure $d\mu(\lambda) =
w(\lambda) \, d\lambda$ and define the define the $n$th correlation
function is given by  
\[
R_n(x_1, x_2, \ldots, x_n) = \frac{1}{Z_N^{\mu} (N-n)!}
\int_{\R^{N-n}} \Omega_N(x_1, \ldots, x_n, y_1, \ldots, y_{N-n}) \,
dy_1 \cdots dy_{N-n}.
\]
It shall be convenient to abbreviate things so that 
\[
\Delta(\bs \lambda) = \prod_{m < n} (\lambda_n - \lambda_m);
\qquad \bs \lambda \in \R^N, 
\]
and if $\mathbf x \in \R^n$ and $\mathbf y \in \R^{N-n}$ then 
\[
\mathbf x \vee \mathbf y = (x_1, \ldots, x_n, y_1, \ldots, y_{N-n}).
\]
Thus,
\[
R_n(\mathbf x) = \frac{1}{Z_N^{\mu} (N-n)!} \int_{\R^{n-n}}
| \Delta(\mathbf x \vee \mathbf y) |^{\beta} \, d\mu^{N-n}(\mathbf y).
\]
The $n$th correlation function is clearly a renormalized version of
the $n$th marginal density of $\Omega_N$.  However, this observation
belies the importance of correlation functions in random matrix
theory, point processes and statistical physics (see for instance
\cite{MR2552864, detpp}).

Given indeterminants $c_1, c_2, \ldots, c_N$ and real numbers $x_1,
x_2, \ldots, x_N$ we define the measure $\nu$ on $\R$ by
\[
d\nu(\lambda) = w(\lambda) \sum_{n=1}^N c_n
d\delta(\lambda - x_n),
\]
where $\delta$ is the probability measure with unit mass at 0.  It is
then a straightforward exercise in symbolic manipulation to show that
$Z_N^{\mu + \nu}/Z_N^{\mu}$ is the generating function for the correlation
functions.  That is, $R_N(x_1, \ldots, x_n)$ is the coefficient of
$c_1 c_2 \cdot \ldots \cdot c_n$ in $Z_N^{\mu + \nu}/Z_N^{\mu}$.

When $\beta = 1$ or 4, this observation and the fact that $Z_N^{\mu +
  \nu}$ is the first step in showing that the correlation functions
have a Pfaffian formulation.  The simplest derivation is given by
Tracy and Widom \cite{MR1657844} using the fact that if $\mathbf A$
and $\mathbf B$ are (perhaps rectangular) matrices for which both
$\mathbf{AB}$ and $\mathbf{BA}$ are square, then 
\begin{equation}
\label{eq:9}
\det(\mathbf I + \mathbf{AB}) = \det(\mathbf I + \mathbf{BA}).
\end{equation}
A similar fact is true for Pfaffians
\cite{rains-2000} which streamlines the proof \cite{borodin-2008,
  sinclair-2008}.  

The existence of a hyperpfaffian representation of $Z_N^{\mu + \nu}$
when $\beta$ is a square is suggestive of a hyperpfaffian formulation
of the correlation functions, however the necessary analog of
(\ref{eq:9}) for hyperpfaffians remains unknown.  

\section{The Proof of Theorem~\ref{thm:1}}

\subsection{Case: $\beta = L^2$ even} 

We define the confluent $NL \times NL$ Vandermonde matrix by first
defining the $NL \times L$ matrix 
\[
\mathbf V(x) = \left[
D^{\ell-1} p_n(x)
\right]_{n,\ell=1}^{NL, L},
\]
and then defining 
\[
\mathbf V(\bs \lambda) = \begin{bmatrix}
 \mathbf V(\lambda_1) &  \mathbf V(\lambda_2) & \cdots &  \mathbf
 V(\lambda_N)
\end{bmatrix}.
\]
The confluent Vandermonde identity has that 
\[
\det \mathbf V(\bs \lambda) = \prod_{m < n} (\lambda_n - \lambda_m)^{L^2} =
\prod_{m < n} | \lambda_n - \lambda_m |^{\beta}.
\]
It follows that
\[
Z_N = \frac{1}{N!}\int_{\R^N} \det \mathbf V(\bs \lambda) \, d\nu^N(\bs \lambda).
\]

We may use Laplace expansion to write $\det \mathbf V(\bs \lambda)$ as a sum
of $N$-fold products of $L \times L$ determinants, where each determinant
appearing in the products is a function of exactly one of the variables
$\lambda_1, \lambda_2, \ldots, \lambda_N$.  That is, given $\mf t: \ul
L \nearrow \ul{NL}$, define $\mathbf V_{\mf t}(\lambda)$ to be the $L
\times L$ matrix,
\[
\mathbf V_{\mf t}(\lambda) = \left[ D^{\ell-1} p_{\mf t(n)}(\lambda)
\right]_{n,\ell=1}^L.
\]
Then, we may write $\det \mathbf V(\bs \lambda)$ as an alternating sum
over products of the form
\[
\det \mathbf V_{\mf t_1}(\lambda_1) \cdot \det \mathbf V_{\mf
  t_2}(\lambda_2) \cdot \ldots \cdot \det \mathbf V_{\mf t_N}(\lambda_N),
\]
where $\mf t_1, \mf t_2, \ldots, \mf t_N: \ul L \nearrow \ul{NL}$ have
ranges which are mutually disjoint (or equivalent, the disjoint union
of their ranges is all of $\ul{NL}$).  The sign of each term in the
sum can be specified by defining $\sgn(\mf t_1, \mf t_2, \ldots, \mf
t_N)$ via 
\[
\epsilon_{\mf t_1} \wedge \epsilon_{\mf t_2} \wedge \cdots \wedge
\epsilon_{\mf t_N} = \sgn(\mf t_1, \mf t_2, \ldots, \mf t_N) \cdot
\epsilon_{\Vol}.
\]
\sloppy{We remark that, if $\mf t_1, \mf t_2, \ldots, \mf t_N: \ul L \nearrow
\ul{NL}$ do not have disjoint ranges, then $\sgn(\mf t_1, \mf t_2, \ldots, \mf
t_N) = 0$.  Thus we can write }
\begin{equation}
\label{eq:3}
\det \mathbf V(\bs \lambda) = \sum_{(\mf t_1, \ldots, \mf t_N)}
\sgn(\mf t_1, \mf t_2, \ldots, \mf t_N)  \det \mathbf V_{\mf t_1}(\lambda_1) \cdot \det \mathbf V_{\mf
  t_2}(\lambda_2) \cdot \ldots \cdot \det \mathbf V_{\mf
  t_N}(\lambda_N), 
\end{equation}
and
\begin{align*}
\int_{\R^N} \det \mathbf V(\bs \lambda) d\nu^N(\bs \lambda) &=
\sum_{(\mf t_1, \ldots, \mf t_N)} 
\sgn(\mf t_1, \mf t_2, \ldots, \mf t_N)  \prod_{n=1}^N \int_{\R} \det
\mathbf V_{\mf t_n}(\lambda) \, d\nu(\lambda) \\
&=
\sum_{(\mf t_1, \ldots, \mf t_N)} 
\sgn(\mf t_1, \mf t_2, \ldots, \mf t_N)  \prod_{n=1}^N \int_{\R}
\Wr(\mathbf p_{\mf t_n}; x) \, d\nu(x) 
\end{align*}
It follows that
\begin{align*}
&\frac{1}{N!} \bigg\{ \int_{\R^N} \det \mathbf V(\bs \lambda)
d\nu^N(\bs \lambda) \bigg\} \epsilon_{\Vol}  \\
& \hspace{2cm} = \frac{1}{N!}\sum_{(\mf
  t_1, \ldots, \mf t_N)} \bigg\{
\prod_{n=1}^N \int_{\R} \Wr(\mathbf p_{\mf t_n}; x) \, d\nu(x)
\bigg\} \sgn(\mf t_1, \mf t_2, \ldots, \mf t_N) \epsilon_{\Vol}  \\
& \hspace{2cm}= \frac{1}{N!}\sum_{(\mf
  t_1, \ldots, \mf t_N)} \bigg\{
\bigwedge_{n=1}^N \int_{\R} \Wr(\mathbf p_{\mf t_n}; x) \, d\nu(x)
\epsilon_{\mf t_n} 
\bigg\}.
\end{align*}
Finally, exchanging the sum and the wedge product, we find 
\begin{align*}
\frac{1}{N!} \bigg\{ \int_{\R^N} \det \mathbf V(\bs \lambda)
d\nu^N(\bs \lambda) \bigg\} \epsilon_{\Vol} &= \frac{1}{N!} \bigg\{ \sum_{\mf t: \ul N \nearrow \ul{NL}} 
\int_{\R} \Wr(\mathbf p_{\mf t}; x) \, d\nu(x)
\epsilon_{\mf t} 
\bigg\}^{\wedge N},
\end{align*}
which proves this case of the theorem.  

\subsection{Case: $\beta = L^2$ odd, $N$ even} 
\label{sec:case:-beta-=}

When $L$ is odd, the situation is complicated by the fact that\[
\prod_{m < n} |\lambda_n - \lambda_m|^{\beta} = \big| \det \mathbf
V(\bs \lambda) \big|.
\]
(Note the absolute values).  Thus, in this case
\[
Z_N = \frac{1}{N!}\int_{\R^N} | \det \mathbf V(\bs \lambda) | \,
d\nu^N(\bs \lambda). 
\]
This complication is eased by the observation,
\cite[Eq.~5.3]{MR0079647}, that if  
\begin{equation}
\mathbf T(\bs \lambda) = \left[ \sgn(\lambda_n - \lambda_m) \right]_{m,n=1}^N,
\qthen \Pf \mathbf T(\bs \lambda) = \prod_{m < n} \sgn(\lambda_n -
\lambda_m),
\label{eq:5}
\end{equation}
and thus
\[
Z_N = \frac{1}{N!}\int_{\R^N} \det \mathbf V(\bs \lambda) \, \Pf \mathbf T(\bs
\lambda) \, d\nu^N(\bs \lambda).
\]
We remark, it is at this point that we require $N$ be even, since the
Pfaffian of $\mathbf T(\bs \lambda)$ is not defined when $N$ is odd.  

Using (\ref{eq:3}), we have 
\[
Z_N = \frac{1}{N!}\sum_{\mf t_1, \mf t_2, \ldots, \mf t_N} \sgn(\mf t_1, \mf t_2,
\ldots, \mf t_N) \int_{\R^N}
\bigg\{ \prod_{n=1}^N \det \mathbf V_{\mf t_n}(\bs \lambda_n) \bigg\}
\Pf \mathbf T(\bs \lambda) \, d\nu^N(\bs \lambda),
\]
where as before the sum is over $N$-tuples of functions $\ul L
\nearrow \ul{NL}$.  Now, if $N = 2M$, then 
\[
\Pf \mathbf T(\bs \lambda) = \frac{1}{M! 2^M} \sum_{\sigma \in S_N}
\sgn \sigma \prod_{m=1}^M \sgn(\lambda_{\sigma(2m)} -
\lambda_{\sigma(2m-1)} ),
\]
and
\[
\prod_{n=1}^N \det \mathbf V_{\mf t_n}(\lambda_n) = \prod_{m=1}^M \det
\mathbf V_{\mf t_{\sigma(2m-1)}}(\lambda_{\sigma(2m-1)}) \det
\mathbf V_{\mf t_{\sigma(2m)}}(\lambda_{\sigma(2m)}).
\]
Thus,
\begin{align*}
Z_N &= \frac{1}{N! M! 2^M}\sum_{\sigma \in S_N} \sum_{\mf t_1, \mf
  t_2, \ldots, \mf t_N} \sgn \sigma \, \sgn(\mf t_1, \mf t_2, \ldots,
\mf t_N)  \\
& \hspace{2cm} \times \int_{\R^N} \bigg\{ \prod_{m=1}^M \det
\mathbf V_{\mf t_{\sigma(2m-1)}}(\lambda_{\sigma(2m-1)}) \det
\mathbf V_{\mf t_{\sigma(2m)}}(\lambda_{\sigma(2m)}) \\
& \hspace{7cm} \times \sgn(\lambda_{\sigma(2m)} -
\lambda_{\sigma(2m-1)} )  \bigg\} d\nu^N(\bs \lambda) \\
& = \frac{1}{N! M!}\sum_{\sigma \in S_N} \sum_{\mf t_1, \mf
  t_2, \ldots, \mf t_N} \sgn \sigma \, \sgn(\mf t_1, \mf t_2, \ldots,
\mf t_N) \\
& \hspace{2cm} \times \bigg\{ \prod_{m=1}^M \frac12 \int_{\R^2} \det
\mathbf V_{\mf t_{\sigma(2m-1)}}(x) \det
\mathbf V_{\mf t_{\sigma(2m)}}(y) \sgn(y - x)
d\nu(x) d\nu(y) \bigg\}.
\end{align*}
Next, we notice that $S_N$ acts on the $N$-tuples $(\mf t_1, \mf t_2,
\ldots, \mf t_N)$ by permutation, and if 
\[
(\mf u_1, \mf u_2, \ldots, \mf u_N) = (\mf t_{\sigma(1)}, \mf
t_{\sigma(2)}, \ldots, \mf t_{\sigma(N)} ),
\]
then
\[
\sgn(\mf u_1, \mf
u_2, \ldots, \mf u_N) = \sgn(\sigma) \sgn(\mf t_1, \mf t_2, \ldots,
\mf t_N).
\]
Moreover, the action of each $\sigma$ produces a bijection on the set
of $N$-tuples, and thus 
\begin{align*}
Z_N &=  \frac{1}{N! M!} \sum_{\sigma \in S_N} \sum_{\mf u_1, \mf u_2,
  \ldots, \mf u_N} \sgn(\mf u_1, \mf u_2, \ldots, \mf u_N) \\
& \hspace{3cm} \times  \bigg\{ \prod_{m=1}^M \int_{\R^2} \frac12 \det
\mathbf V_{\mf u_{2m-1}}(x) \det \mathbf V_{\mf u_{2m}}(y)  \sgn(y - x)
d\nu(x) d\nu(y) \bigg\}. 
\end{align*}
The summand is independent of $\sigma$, and thus 
\begin{align*}
Z_N &=  \frac{1}{M!} \sum_{\mf u_1, \mf u_2,
  \ldots, \mf u_N} \sgn(\mf u_1, \mf u_2, \ldots, \mf u_N) \\
& \hspace{2cm} \times \bigg\{
\prod_{m=1}^M \frac12 \int_{\R^2} \det 
\mathbf V_{\mf u_{2m-1}}(x) \det \mathbf V_{\mf u_{2m}}(y)  \sgn(y - x)
d\nu(x) d\nu(y) \bigg\}. 
\end{align*}
Now, every $N$-tuple $(\mf u_1, \mf u_2, \ldots, \mf u_N)$ can be
obtained from an $M$-tuple $(\mf v_1, \mf v_2, \ldots, \mf v_M)$ of
functions $\ul{2L} \nearrow \ul{2LM}$ and an $M$-tuple $(\mf w_1, \mf
w_2, \ldots, \mf w_M)$ of functions $\ul L \nearrow \ul{2L}$ specified
by  
\[
\mf u_{2m-1} = \mf v_m \circ \mf w_m \qand \mf u_{2m} = \mf v_m \circ
\mf w_m',
\]
where $\mf w_m'$ is the unique function $\ul L \nearrow \ul{2L}$ whose range
is disjoint from $\mf w_m$.  Defining the sign of each of the $\mf
w_m$ by 
\[
\epsilon_{\mf w_m} \wedge \epsilon_{\mf w_m'} = \sgn \mf w_m (\mathbf
e_1 \wedge \mathbf e_2 \wedge \cdots \wedge \mathbf e_{2L} ),
\]
we have 
\[
\sgn(\mf u_1, \mf u_2, \ldots, \mf u_N) = \sgn(\mf v_1, \mf v_2,
\ldots, \mf v_M) \prod_{m=1}^M \sgn \mf w_m.
\]
Using this decomposition,
\begin{align*}
  Z_N &= \frac{1}{M!} \sum_{\mf v_1, \mf v_2, \ldots, \mf v_M}
  \sgn(\mf v_1, \mf v_2, \ldots, \mf v_M)
 \\
  & \quad \times   \sum_{\mf w_1, \mf w_2, \ldots, \mf
    w_m} \! \! \!\bigg\{ \prod_{m=1}^M \frac{\sgn \mf w_m}2
  \int_{\R^2} \det \mathbf V_{\mf v_m \circ \mf w_m}(x) \det \mathbf
  V_{\mf v_m \circ \mf w_m'}(y) \sgn(y - x)
  d\nu(x) d\nu(y) \bigg\} \\
  &= \frac{1}{M!} \sum_{\mf v_1, \mf v_2, \ldots, \mf v_M} \sgn(\mf
  v_1, \mf v_2, \ldots, \mf v_M) \\
  & \quad \times \bigg\{ \prod_{m=1}^M \bigg[ \frac12
  \sum_{\mf w: \ul L \nearrow \ul{2L}} \sgn \mf w \int_{\R^2} \det
  \mathbf V_{\mf v_m \circ \mf w}(x) \det \mathbf
  V_{\mf v_m \circ \mf w'}(y) \sgn(y - x) d\nu(x) d\nu(y) \bigg]
  \bigg\},
\end{align*}
and consequently, 
\begin{align*}
Z_N \epsilon_{\vol} & = \frac{1}{M!} \sum_{\mf v_1, \mf v_2, \ldots, \mf v_M} \bigg\{
\bigwedge_{m=1}^M  \\ & \hspace{1cm} \bigg[ \frac12 \sum_{\mf w: \ul L 
   \nearrow \ul{2L}} \sgn \mf w \int_{\R^2} \det  
\mathbf V_{\mf v_m \circ \mf w}(x) \det \mathbf V_{\mf v_m \circ \mf
w'}(y)  \sgn(y - x) 
d\nu(x) d\nu(y) \bigg] \epsilon_{\mf v_m} \bigg\} \\
& = \frac{1}{M!} \bigg(\frac12 \sum_{\mf v: \ul{2L} \nearrow \ul{2LM}}
\\ & \hspace{1cm} \bigg[
  \sum_{\mf w: \ul L  \nearrow \ul{2L}} \sgn \mf w \int_{\R^2} \det   
\mathbf V_{\mf v \circ \mf w}(x) \det \mathbf V_{\mf v \circ \mf
w'}(y) \sgn(y - x) 
d\nu(x) d\nu(y) \bigg] \epsilon_{\mf v}  \bigg)^{\wedge M}.
\end{align*}
Thus, substituting $\Wr(\mathbf p_{\mf v \circ \mf w}; x) = \det
\mathbf V_{\mf v \circ \mf w}(x)$ (and likewise for $\Wr(\mathbf
p_{\mf v \circ \mf w'}; y)$) we have $Z_N$ equal to 
\[
\PF \bigg(\frac12 \sum_{\mf v: \ul{2L} \nearrow \ul{2LM}} \bigg[
  \sum_{\mf w: \ul L  \nearrow \ul{2L}} \sgn \mf w \int_{\R^2} 
\Wr(\mathbf p_{\mf v \circ \mf w}; x) \Wr(\mathbf p_{\mf v \circ \mf
w'};y) \sgn(y - x) 
d\nu(x) d\nu(y) \bigg] \epsilon_{\mf v}  \bigg).
\]
Next, we use the fact that $\epsilon_{\mf v} = \sgn \mf w \,
\epsilon_{\mf v \circ \mf w} \wedge \epsilon_{\mf v \circ \mf w'}$ to   
write 
\begin{align}
&\frac{1}{2}\sum_{\mf v: \ul{2L} \nearrow \ul{2LM}} \bigg[
  \sum_{\mf w: \ul L  \nearrow \ul{2L}} \sgn \mf w  \int_{\R^2} 
\Wr(\mathbf p_{\mf v \circ \mf w}; x) \Wr(\mathbf p_{\mf v \circ \mf
w'};y)  \sgn(y - x) 
d\nu(x) d\nu(y) \bigg] \epsilon_{\mf v} \nonumber \\
&  = 
\frac12 \sum_{\mf v: \ul{2L} \nearrow \ul{2LM}} 
  \sum_{\mf w: \ul L  \nearrow \ul{2L}} \bigg\{ \int_{\R^2} 
\Wr(\mathbf p_{\mf v \circ \mf w}; x) \Wr(\mathbf p_{\mf v \circ \mf
w'};y)  \sgn(y - x)
d\nu(x) d\nu(y) \bigg\} \epsilon_{\mf v \circ \mf w} \wedge \epsilon_{\mf
  v \circ \mf w'}. \label{eq:13}
\end{align}
Finally, we see that as $\mf v$ varies over $\ul{2L} \nearrow
\ul{2LM}$ and $\mf w$ varies over  $\ul L  \nearrow \ul{2L}$, $\mf t =
\mf v \circ \mf w$ and $\mf u = \mf v \circ \mf w'$ vary over pairs
in $\ul L \nearrow \ul{2LM}$ with disjoint ranges.  However, since
$\epsilon_{\mf t} \wedge \epsilon_{\mf u} = 0$ if the ranges of $\mf
t$ and $\mf u$ are not disjoint, we may replace the double sum in
(\ref{eq:13}) with a double sum over $\ul L \nearrow \ul{2LM}$. That
is, (\ref{eq:13}) equals
\begin{align*}
&  
\frac12 \sum_{\mf t: \ul L \nearrow \ul{2LM}} \sum_{\mf u: \ul L \nearrow
  \ul{2LM}} \bigg\{ \int_{\R^2} 
\Wr(\mathbf p_{\mf t}; x) \Wr(\mathbf p_{\mf u};y)  \sgn(y - x)
d\nu(x) d\nu(y) \bigg\} \epsilon_{\mf t} \wedge \epsilon_{\mf u} \\
& \qquad = 
\frac{1}{2}  \int_{\R^2}  \bigg\{ \sum_{\mf t: \ul L \nearrow
  \ul{2LM}} \sum_{\mf u: \ul L \nearrow  \ul{2LM}} 
\Wr(\mathbf p_{\mf t}; x) \Wr(\mathbf p_{\mf u};y) \sgn(y - x)
 \epsilon_{\mf t} \wedge \epsilon_{\mf u}  \bigg\} d\nu(x) d\nu(y) \\
& \qquad = \frac{1}{2}  \int_{\R^2}  \omega(x) \wedge \omega(y)
\sgn(y - x)
 d\nu(x) d\nu(y),
\end{align*}
as desired.  

\subsection{Case: $\beta = L^2$ odd, $N$ odd}

In this section we will assume that $N = 2K - 1$ is odd and for $\bs
\lambda \in \R^N$ we will introduce the $2K \times 2K$ matrix $\mathbf
T'(\bs \lambda)$ by
\[
\mathbf T;(\bs \lambda) = \big[ t_{m,n} \big] = \begin{bmatrix}
& & & 1 \\
& \mathbf T(\bs\lambda)& & \vdots \\
& & & 1 \\
-1 & \cdots & -1 & 0
\end{bmatrix}.
\]
That is 
\[
\mathbf T'(\bs \lambda) =  \mathbf T(\bs \lambda
\vee \infty) = \lim_{t 
  \rightarrow \infty} \mathbf T(\bs \lambda \vee t),
\]
where, for instance, $(\bs \lambda \vee t) = (\lambda_1, \ldots,
\lambda_N, t) \in \R^{N+1}$.  From (\ref{eq:5}), 
\[
\PF \mathbf T'(\bs \lambda) = \prod_{1 \leq m < n \leq N}
\sgn(\lambda_n - \lambda_m). 
\]
Here we will write 
\[
\Pi_{2K} = \{ \sigma \in S_{2K} : \sigma(2k-1) < \sigma(2k) \mbox{ for
} k=1,2,\ldots,K \},
\]
and expand the Pfaffian of $\mathbf U(\bs \lambda)$ as 
\[
\PF \mathbf U(\bs \lambda) = \frac{1}{K!} \sum_{\sigma \in \Pi_{2K}}
\sgn \sigma \prod_{k=1}^K t_{\sigma(2k-1), \sigma(2k)}.
\]
For each $\sigma \in \Pi_{2K}$ there exists $k_{\sigma} \in \ul{K}$
such that $\sigma(2 k_{\sigma}) = 2K$, and hence,
$t_{\sigma(2k_{\sigma} - 1), \sigma(2k_{\sigma})} = 1$.  That is,
\[
\PF \mathbf U(\bs \lambda) = \frac{1}{K!} \sum_{\sigma \in \Pi_{2K}}
\sgn \sigma
\prod_{k=1 \atop k \neq k_{\sigma}}^K \sgn(\lambda_{\sigma(2k)} -
\lambda_{\sigma(2k-1)}).
\]
Using this, and following the outline of the $L$ odd $N$ even case, we
find
\begin{align*}
Z_N &= \frac{1}{N!K!} \sum_{\sigma \in \Pi_{2K}} \sum_{\mf t_1, \mf
  t_2, \ldots, \mf t_N} \sgn \sigma \, \sgn(\mf t_1, \mf t_2, \ldots,
\mf t_N)   \int_{\R^N} \Wr(\mathbf p_{\mf
  t_{\sigma(2k_{\sigma} - 1)}}; \lambda_{\sigma(2k_{\sigma}-1)})
 \\
& \qquad \times \bigg\{ \prod_{k=1 \atop k \neq
  k_{\sigma}}^K \Wr(\mathbf p_{\mf t_{\sigma(2k-1)}};
\lambda_{\sigma(2k-1)}) \Wr(\mathbf p_{\mf t_{\sigma(2k)}}; \lambda_{\sigma(2k)}) \sgn(\lambda_{\sigma(2k)} -
\lambda_{\sigma(2k-1)} )  \bigg\}   d\nu^N(\bs \lambda),
\end{align*}
where, as before, $(\mf t_1, \mf t_2, \ldots, \mf t_N)$ is an
$N$-tuple of functions $\ul L \nearrow \ul{NL}$.  Fubini's Theorem
then yields
\begin{align*}
Z_N &= \frac{1}{N!K!} \sum_{\sigma \in \Pi_{2K}} \sum_{\mf t_1, \mf
  t_2, \ldots, \mf t_N} \sgn \sigma \, \sgn(\mf t_1, \mf t_2, \ldots,
\mf t_N)  \int_{\R} \Wr(\mathbf p_{\mf
  t_{\sigma(2k_{\sigma} - 1)}}; x) d\nu(x) \\
& \hspace{2cm}  \times  \bigg\{ \prod_{k=1 \atop k \neq
  k_{\sigma}}^K\int_{\R^2} \Wr(\mathbf p_{\mf
  t_{\sigma(2k-1)}}; x) \Wr(\mathbf p_{\mf t_{\sigma(2k)}}; y) \sgn(y -
x) \, d\nu(x) d\nu(y) \bigg\} 
\end{align*}
Now, given $\sigma \in \Pi_{2K}$, let $\wt \sigma \in \Pi_{2K}$ be the permutation
given by
\[
\wt \sigma(n) = \left\{
\begin{array}{ll}
\sigma(n) & \mbox{ if  } n \leq 2k_{\sigma}; \\
\sigma(n+2) & \mbox{ if  } 2k_{\sigma} + 2 \leq n \leq 2K-2; \\
\sigma(2k_{\sigma}-1)  & \mbox{ if  } n = 2K - 1; \\
2K  & \mbox{ if  } n = 2K. \\
\end{array}
\right.
\]
That is, $\wt \sigma$ is the permutation whose range as an ordered set
is equal to that formed from the range of $\sigma$ by `moving' 
$\sigma(2 k_{\sigma} - 1)$ and $\sigma(2 k_{\sigma})$ to the `end.'
Clearly $\sgn \wt \sigma = \sgn \sigma$, and each $\wt \sigma$
corresponds to exactly $K$ distinct $\sigma \in \Pi_{2K}$.  It follows that\begin{align*}
Z_N &= \frac{1}{N!K!} \sum_{\sigma \in \Pi_{2K}} \sum_{\mf t_1, \mf
  t_2, \ldots, \mf t_N} \sgn \wt \sigma \, \sgn(\mf t_1, \mf t_2, \ldots,
\mf t_N)  \int_{\R} \Wr(\mathbf p_{\mf
  t_{\wt \sigma(2K - 1)}}; x) d\nu(x) \\
& \hspace{2cm}  \times  \bigg\{ \prod_{k=1}^{K-1}\int_{\R^2} \Wr(\mathbf p_{\mf
  t_{\wt \sigma(2k-1)}}; x) \Wr(\mathbf p_{\mf t_{\wt \sigma(2k)}}; y) \sgn(y -
x) \, d\nu(x) d\nu(y) \bigg\} \\ 
\end{align*}
Next, denote the transpositions $\tau_k = (2k-1, 2k)$ for $k = 1, 2,
\ldots, K-1$ and let $G_{2K-2}$ be the group of permutations generated
by the $\tau_k$.  Clearly the cardinality of $G_{2K-2}$ is $2^{K-1}$,
and moreover the map 
\begin{equation}
\label{eq:15}
\Pi_{2K} \times G_{2K-2} \rightarrow \{ \pi \in S_{2K} : \pi(2K) = 2K
\} 
\end{equation}
given by $(\sigma, \tau) \mapsto \wt\sigma \circ \tau$ is a $K$ to one
map.  Clearly, the right hand set is in correspondence with $S_N$.  
Now, if $\pi = \wt \sigma \circ \tau$ for some $\sigma \in \Pi_{2K}$ and $\tau \in G_{2K-2}$ then,
\begin{align*}
& \sgn \wt \sigma \bigg\{\!\! \prod_{k=1}^{K-1}\int\limits_{\R^2}
\Wr(\mathbf p_{\mf t_{\wt \sigma(2k-1)}}; x) \Wr(\mathbf p_{\mf t_{\wt
    \sigma(2k)}}; y) \sgn(y - 
x) \, d\nu(x) d\nu(y) \bigg\} 
\int\limits_{\R} \Wr(\mathbf p_{\mf
  t_{\wt \sigma(2K - 1)}}; x) d\nu(x) \\
&\hspace{2cm} = \sgn \pi \bigg\{ \prod_{k=1}^{K-1}\int\limits_{\R^2} \Wr(\mathbf p_{\mf
  t_{\pi(2k-1)}}; x) \Wr(\mathbf p_{\mf t_{\pi(2k)}}; y) \sgn(y -
x) \, d\nu(x) d\nu(y) \bigg\} \\
& \hspace{9cm} \times 
\int\limits_{\R} \Wr(\mathbf p_{\mf
  t_{\pi(2K - 1)}}; x) d\nu(x),
\end{align*}
the factors of $-1$ introduced into $\pi$ by the transpositions in
$\tau$ being compensated by the fact that the double integral
swaps sign when the arguments are swapped.  It follows that we may
replace the sum over $\Pi_{2K}$ with a sum over $S_N$ so long as we
compensate for the cardinality of $G_{2K-2}$ and the fact that the map
is $K$ to 1.  That is, 
\begin{align*}
Z_N &= \frac{1}{N!(K-1)!} \sum_{\pi \in S_N}
\sum_{\mf t_1, \mf 
  t_2, \ldots, \mf t_N} \sgn \pi \, \sgn(\mf t_1, \mf t_2, \ldots,
\mf t_N)  \\
& \hspace{2cm}  \times  \bigg\{ \prod_{k=1}^{K-1} \frac{1}{2}\int_{\R^2}
\Wr(\mathbf p_{\mf t_{\pi(2k-1)}}; x) \Wr(\mathbf p_{\mf
  t_{\pi(2k)}}; y) \sgn(y - x) \, d\nu(x) d\nu(y) \bigg\} \\
& \hspace{9cm} \times \int_{\R} 
\Wr(\mathbf p_{\mf t_{\pi(N)}}; x) d\nu(x).
\end{align*}
As before 
\[
\sgn \pi \sgn(\mf t_1, \mf t_2, \ldots, \mf t_N) = \sgn(
\mf t_{\pi(1)}, \mf t_{\pi(2)}, \ldots, \mf t_{\pi(N)}),
\]
and the action of any particular $\pi \in S_N$ on the $N$-tuple $(\mf
t_1, \mf t_2, \ldots, \mf t_N)$ is a bijection on the set of such
$N$-tuples.  Thus, we may eliminate the sum over $S_N$ so long as we
compensate by $N!$.  That is, 
\begin{align}
Z_N &= \frac{1}{(K-1)!} 
\sum_{\mf t_1, \mf 
  t_2, \ldots, \mf t_N} \sgn(\mf t_1, \mf t_2, \ldots,
\mf t_N)  \label{eq:7} \\
& \hspace{2cm}  \times  \bigg\{ \prod_{k=1}^{K-1} \frac{1}{2} \int_{\R^2}
\Wr(\mathbf p_{\mf t_{2k-1}}; x) \Wr(\mathbf p_{\mf t_{2k}}; y)
\sgn(y - x) \, d\nu(x) d\nu(y) \bigg\} \nonumber \\
& \hspace{9cm} \times \int_{\R} \Wr(\mathbf p_{\mf
  t_{N}}; x) d\nu(x). \nonumber
\end{align}

\sloppy{With a slight modification of the argument in the $L$ odd $N$ even
case, we set $\mf t = \mf t_N$ and write $(\mf t_1, \mf t_2, \ldots, \mf t)$ as}
\begin{equation}
\label{eq:6}
(\mf t' \circ \mf v_1 \circ \mf w_1, \mf t' \circ \mf v_1 \circ
\mf w_1', \ldots, \mf t' \circ \mf v_{K-1} \circ \mf w_{K-1}, \mf
t' \circ \mf v_{K-1} \circ \mf w_{K-1}', \mf t),
\end{equation}
where $(\mf v_1, \mf v_2, \ldots, \mf v_{K-1})$ is a $K-1$-tuple of
functions $\ul{2L} \nearrow \ul{(N-1)L}$ the union of ranges of which
is all of $\ul{L(N-1)}$, and $(\mf w_1, \mf w_2, \ldots, \mf w_{K-1})$
is a $K-1$-tuple of functions $\ul L \nearrow \ul{2L}$.  In words,
(\ref{eq:6}) is formed by first forming a partition of $\ul{(N-1)L}$
into $L$ disjoint subsets given by the ranges of the
$\mf v$s.  The $\mf w$s serve to divide each of these sets
into two equal sized sets giving a partition of $\ul{(N-1)L}$ into
$2L$ disjoint subsets.  Finally, $\mf t'$ is the function from
$\ul{N(L-1)} \nearrow \ul{NL}$ whose range is disjoint from $\mf
t_N$.  Each $(\mf t_1, \mf t_2, \ldots, \mf t_N)$ has a unique
representation of this form, and 
\[
\sgn(\mf t_1, \mf t_2, \ldots, \mf t_{N-1}, \mf t) = \sgn \mf t' \, \sgn(\mf v_1,
\mf v_2, \ldots, \mf v_{K-1}) \prod_{k=1}^{K-1} \sgn \mf w_k. 
\]

With these observations, we may rewrite (\ref{eq:7}) as
\begin{align*}
Z_N &= \frac{1}{(K-1)!} \sum_{\mf t: \ul L \nearrow \ul{NL}}  \bigg[ \sgn \mf t'
\int_{\R} \Wr(\mathbf p_{\mf  t}; x) d\nu(x) \sum_{\mf v_1,
 \ldots, \mf v_{K-1}}  \sum_{\mf w_1, \ldots, \mf w_{K-1}} \!\!\! \sgn(\mf
v_1, \mf v_2, \ldots, \mf v_{K-1}) \\
& \hspace{1cm}  \times \bigg\{ \prod_{k=1}^{K-1}  \frac{\sgn \mf w_k}{2}
\int_{\R^2} \Wr( \mathbf p_{\mf t'  \circ \mf v_k \circ \mf w_k}; x)
\Wr(\mathbf p_{\mf t' \circ \mf v_k 
  \circ \mf w_k'}; y) \sgn(y - x) \, d\nu(x) d\nu(y) \bigg\} \bigg],
\end{align*}
and thus, since $\epsilon_{\vol} = \sgn \mf t' \; \epsilon_{\mf t'} \wedge
\epsilon_{\mf t}$, 
\begin{align*}
 &Z_N \epsilon_{\vol} = \frac{1}{(K-1)!} \sum_{\mf t: \ul L \nearrow \ul {NL}}
\bigg[ \int_{\R} \Wr(\mathbf p_{\mf t}; x) d\nu(x) 
\sum_{\mf v_1,  \ldots, \mf v_{K-1}}
\sum_{\mf w_1, \ldots, \mf w_{K-1}} \sgn(\mf v_1, \mf 
v_2, \ldots, \mf v_{K-1})\\
& \qquad   \times \prod_{k=1}^{K-1} \bigg\{
 \frac{\sgn \mf w_k}{2} \int_{\R^2} \Wr(\mathbf p_{\mf t'
  \circ \mf v_k \circ \mf w_k}; x) \Wr(\mathbf p_{\mf t' \circ \mf v_k
  \circ \mf w_k'}; y) \sgn(y - x) \, d\nu(x) d\nu(y) \bigg\}
\epsilon_{\mf t'} \wedge \epsilon_{\mf t} \bigg].
\end{align*}
Now, 
\[
\epsilon_{\mf t'} = \sgn(\mf v_1, \mf v_2, \ldots, \mf v_{K-1})
\bigg\{ \prod_{k=1}^{K-1} \sgn \mf w_k  \bigg\} \bigwedge_{k=1}^{K-1}
\epsilon_{\mf t' \circ \mf v_k \circ \mf w_k} \wedge \epsilon_{\mf t'
  \circ \mf v_k \circ \mf w'_k} 
\]
and thus,
\begin{align*}
 &Z_N \epsilon_{\vol} = \frac{1}{(K-1)!} \sum_{\mf t: \ul L \nearrow \ul {NL}}
\bigg[ \sum_{\mf v_1,  \ldots, \mf v_{K-1}}
\sum_{\mf w_1, \ldots, \mf w_{K-1}} \\
& \hspace{2cm} \bigg\{ \bigwedge_{k=1}^{K-1}
 \frac{1}{2} \int_{\R^2} \Wr(\mathbf p_{\mf t'
  \circ \mf v_k \circ \mf w_k}; x) \Wr(\mathbf p_{\mf t' \circ \mf v_k
  \circ \mf w_k'}; y) \\
& \hspace{6cm} \times \sgn(y - x) \, d\nu(x) d\nu(y) \; \epsilon_{\mf
  t' \circ \mf v_k \circ \mf w_k} \wedge \epsilon_{\mf t' \circ \mf
  v_k \circ \mf w'_k} \bigg\} \\
& \hspace{10cm} \wedge  \int_{\R} \Wr(\mathbf p_{\mf t}; x) d\nu(x) \;
\epsilon_{\mf t} \bigg].
\end{align*}
Using more-or-less the same maneuvers as in the $L$ odd, $N$ even
case, we may exchange the sums and the $K-1$-fold wedge product to find
\begin{align*}
& Z_N \epsilon_{\vol} =\sum_{\mf t: \ul L \nearrow \ul {NL}}
 \frac{1}{(K-1)!} \\
& \hspace{.5cm}  \bigg( \sum_{\mf s,\mf u: \ul{L} \nearrow
  \ul{(N-1)L}}  \frac12 \int_{\R^2} \Wr(\mathbf p_{\mf t' \circ
  \mf s}; x) \Wr(\mathbf p_{\mf   t' \circ \mf u}; y) 
\sgn(y -   x) \, d\nu(x) d\nu(y) \; \epsilon_{\mf t'\circ \mf s}
\wedge \epsilon_{\mf t'\circ \mf u} 
\bigg)^{\wedge(K-1)} \\
& \hspace{10cm} \wedge \int_{\R}
\Wr(\mathbf p_{\mf t}; x) d\nu(x) \; \epsilon_{\mf t},
\end{align*}
and thus, 
\begin{align*}
& Z_N \epsilon'_{\vol} =\sum_{\mf t: \ul L \nearrow \ul{NL}}
 \frac{1}{(K-1)!} \\
& \hspace{.5cm}  \bigg( \sum_{\mf s,\mf u: \ul{L} \nearrow
  \ul{(N-1)L}}  \frac12 \int_{\R^2} \Wr(\mathbf p_{\mf t' \circ
  \mf s}; x) \Wr(\mathbf p_{\mf   t' \circ \mf u}; y) 
\sgn(y -   x) \, d\nu(x) d\nu(y) \; \epsilon_{\mf t'\circ \mf s}
\wedge \epsilon_{\mf t'\circ \mf u} 
\bigg)^{\wedge(K-1)} \\
& \hspace{9.5cm} \wedge \int_{\R}
\Wr(\mathbf p_{\mf t}; x) d\nu(x) \; \epsilon_{\mf t} \wedge \epsilon',
\end{align*}
Next, we notice that we may extend the sum in the $K-1$-wedge product
to all basic forms of the form $\epsilon_{\mf s} \wedge \epsilon_{\mf
  u}$, since if the range of $\mf s$ or $\mf u$ has nontrivial
intersection with that of $\mf t$, wedging by $\epsilon_{\mf t} \wedge
\epsilon'$ will cause that term to vanish is the final expression.
That is, 
\begin{align*}
Z_N \epsilon'_{\vol} &=\sum_{\mf t: \ul L \nearrow \ul{NL}}
 \frac{1}{(K-1)!} \\
& \hspace{1.5cm}  \bigg( \sum_{\mf s,\mf u: \ul{L} \nearrow
  \ul{NL}}  \frac12 \int_{\R^2} \Wr(\mathbf p_{\mf s}; x)
\Wr(\mathbf p_{\mf u}; y)  \sgn(y -   x) \, d\nu(x) d\nu(y) \;
\epsilon_{\mf s} \wedge \epsilon_{\mf u} 
\bigg)^{\wedge(K-1)} \\
& \hspace{8cm} \wedge \int_{\R}
\Wr(\mathbf p_{\mf t}; x) d\nu(x) \; \epsilon_{\mf t} \wedge
\epsilon'.
\end{align*}
Now, since the $K-1$-fold wedge product is independent of $\mf t$ we
may factor it out of the sum to write $Z_N \epsilon'_{\vol}$ as 
\begin{align*}
& \frac{1}{(K-1)!}\bigg( \sum_{\mf s,\mf u: \ul{L} \nearrow
  \ul{NL}}  \frac12 \int_{\R^2} \Wr(\mathbf p_{\mf s}; x)
\Wr(\mathbf p_{\mf u}; y)  \sgn(y -   x) \, d\nu(x) d\nu(y) \;
\epsilon_{\mf s} \wedge \epsilon_{\mf u} 
\bigg)^{\wedge(K-1)} \\
& \hspace{7cm} \wedge \sum_{\mf t: \ul L \nearrow \ul{NL}} \int_{\R}
\Wr(\mathbf p_{\mf t}; x) d\nu(x) \; \epsilon_{\mf t} \wedge
\epsilon'. \\
&= \frac{1}{(K-1)!} \bigg( \frac12 \int_{\R^2} \omega(x) \wedge
\omega(y) \, \sgn(y-x) \, d\nu(x) \, d\nu(y) \bigg)^{\wedge(K-1)}
\wedge \int_{\R} \omega(x) \wedge \epsilon' \, d\nu(x),
\end{align*}
where the last equation follows from the definition of $\omega$.  

Finally, using the binomial theorem,
\begin{align*}
&\frac{1}{K!} \bigg(\frac12 \int_{\R} \int_{\R} \omega(x) \wedge
\omega(y) \sgn(y-x) \, d\nu(x) \, d\nu(y) +
\int_{\R} \omega(x) \wedge \epsilon' \, d\nu(x) \bigg)^{\wedge K} \\
& \hspace{2cm}  = \sum_{k=0}^K \frac{1}{(K-k)! k!} \bigg( \frac12
\int\limits_{\R^2} \omega(x) \wedge \omega(y) \, \sgn(y-x) \, d\nu(x)
\, d\nu(y) \bigg)^{\wedge(K-k)} \\
& \hspace{9cm} \wedge \bigg( \int\limits_{\R}
\omega(x) \wedge \epsilon' \, d\nu(x)  \bigg)^{\wedge k}.
\end{align*}
However, since every term in this sum must contain exactly one factor
of $\epsilon'$ we see that the terms corresponding to $k=1$ are the
only ones which contribute.  That is,
\begin{align*}
&\frac{1}{K!} \bigg(\frac12 \int_{\R} \int_{\R} \omega(x) \wedge
\omega(y) \sgn(y-x) \, d\nu(x) \, d\nu(y) +
\int_{\R} \omega(x) \wedge \epsilon' \, d\nu(x) \bigg)^{\wedge K} \\
& \hspace{1cm} = \frac{1}{(K-1)!} \bigg( \frac12 \int_{\R^2} \omega(x) \wedge
\omega(y) \, \sgn(y-x) \, d\nu(x) \, d\nu(y) \bigg)^{\wedge(K-1)}
\wedge \int_{\R} \omega(x) \wedge \epsilon' \, d\nu(x),
\end{align*}
which completes the proof of this case.  

\subsection{Case: $\beta = L^2 + 1$ even, $N$ even}

We use the fact that, if $N = 2M$ and 
\[
\mathbf S(\bs \lambda) = \left[ \frac{(\lambda^M_n -
    \lambda^M_m)^2}{\lambda_n - \lambda_m} \right]_{m,n=1}^{N} \qthen 
\Pf \mathbf S(\bs \lambda) = \prod_{m < n} (\lambda_n - \lambda_m).
\]
(See \cite[Lemma 5.2]{MR2219947}).  It follows that
\begin{align*}
Z_N & = 
\frac{1}{N!} \int_{\R^N} \bigg\{\prod_{m < n} (\lambda_n -
  \lambda_m)^{L^2}\bigg\} \bigg\{\prod_{m < n} (\lambda_n -
  \lambda_m)\bigg\}  \, d\nu^N(\bs \lambda) \\
& = \frac{1}{N!} \int_{\R^N} \det \mathbf V(\bs \lambda) \Pf \mathbf
S(\bs \lambda) \, d\nu^N(\bs \lambda).
\end{align*}
The proof in this case is then the same as given in
Section~\ref{sec:case:-beta-=} by replacing the particulars of the matrix 
$\mathbf T(\bs \lambda)$ with those of the matrix $\mathbf S(\bs \lambda)$.  
\bibliography{bibliography}

\begin{center}
\noindent\rule{4cm}{.5pt}
\vspace{.25cm}

\noindent {\sc \small Christopher D.~Sinclair}\\
{\small Department of Mathematics, University of Oregon, Eugene OR 97403} \\
email: {\tt csinclai@uoregon.edu}
\end{center}

\end{document}